# The Field Structure of Free Photons

Anthony Rizzi
Institute for Advanced Physics, arizzi@iapweb.org

**Abstract:** Using a quantum field theoretic description of the photon it is shown that, as intuitively expected but not before theoretically proven, the vector potential of a photon has a likely amplitude associated with a discrete frequency and therefore energy, and momentum. In particular, by finding the wave-functional for the vector potential, it is shown that the likely absolute amplitude spectrum has delta function at a given frequency. This analysis is extended to *n*-photon systems. It shows that such systems have a vector potential distribution whose most likely element has a strong sinusoidal component which has an amplitude corresponding to *n*-fold more energy than a single photon system. An analogous result for photons of different energy is also derived. Through the use of Parseval's theorem for stochastic systems, the calculations and associated analyses introduces a simple tool for exploring the nature of QFT Schrödinger wave-functional generally.

## Introduction

Photons are discussed everywhere from middle school, even grade school, to graduate textbooks and beyond and are a core part of modern physics, but proving and show their field structure, the focus of this article, has not been a part of these discussions. Rules of thumb and intuitions from ordinary quantum mechanics and quantum optics abound. But, as we know, these do not substitute for rigorous analysis from first principles from what we know theoretically. And, what we know theoretically about photons is contained in quantum field theory. Ordinary quantum mechanics and quantum optics studies can only handle photons to the extent information is imported from QFT. This paper for the first time directly applies the relevant QFT to obtain the full probabilistic structure of the photon's vector potential (which is what specifies what it is) towards proving its key property, its discrete energy nature. In the process, a new technique for probing such structures is introduced. Though QFT is a mature part of physics, no one has yet shown the nature of a photon using QFT in the primal way done here. Seeing this structure and how it connects to our previous intuitions is an essential insight into the physics of photons, which, in turn, are at the center of modern physics.

Much previous work has been done in trying to probe the nature of the photon,[1] beginning with the early investigations trying to "split" the photon.[2] Free photons are often discussed as if they were localized particles, but they are not. Of course, photons by definition cause localized detections, but, as this article, proves their most likely field structure is, in the idealized limit, an infinite length sine wave of undetermined phase. This agrees with the natural understanding of a photon as a packet of *definite* energy (and thus definite frequency), but makes it clear that even though when it interacts, it interacts locally (e.g., with an atom), when it is free, before it begins to interact (with an atom for example), its wave structure is spread out, not a localized wave packet.[3] Giving (via proofs) the structure of the photon offers us insights into simplest QFT structure, the free photon, and thus can be a first glimpse of a deeper view of the nature of QFT itself as well as a formalisms that has seldom been used, the Schrödinger wavefunctional and a statistical formalism that has not been yet applied in QFT.

---

[1] E.G., see special issue on nature of photon: *Optics and Photonics* News (vol. 14, October 2003).
[2] E. O. Lawrence, J. W. Beams, *On the Nature of Light*, PNAS **13** No. 4, 207-212 (1927), G. P. Thomson, *Test of a Theory of Radiation*, Proceedings of the Royal Society of London. Series A, Containing Papers of a Mathematical and Physical Character **104** No. 724, 115-120 (1923).

Up to this point, photon structure has been, indeed, taken for granted as issues related to a wide range of important problems were discussed. These extend from trying to find analogs of ordinary of wavefunctions for the spatially limited photon as well as its associated probability for locating a photon, separately developing probability distributions for detecting a photon based on an electric field operator, finding energy distribution functions of such spatially limited photons to studying correlation functions and coherent states created as superposition of Fock states and photon interference phenomena as well as light squeezing.[4] These problems are addressed using various approaches, but seldom, if ever, use the QFT formalism of the Schrödinger wavefunctional. The most natural approach to the photon field structure is this Schrödinger formalism in combination with the commonly used Fock formalism. Now, each formalism facilitates a different insight into the physics of a given system. Feynman once said that every theorist should know many ways to solve the same problem.[5] The Schrödinger wavefunctional formalism has been somewhat neglected in this regard. In ordinary quantum mechanics, the Schrödinger equation is usually preferentially used in both teaching and applications. In ordinary QM, the *S*-matrix type approach and other approaches (e.g. Green functions and path integrals) have their place and are used, but do not eclipse the Schrödinger equation. By contrast in QFT teaching and applications, the analogous Schrödinger wave functional equation is left out almost completely (along with all the intuitive advantages it affords). Few resources use or even mention it.[6,7] Because of this, we first review those little known results of the analysis for our case and then move to understand what they imply. Deriving those results is helpful in clearly

---

[3] This is sometimes alluded to but usually vaguely but, in the literature as far as I could find, always without proof.

[4] See for example R. Loudon, *The Quantum Theory of Light Third Edition* (Oxford University Press, New York, 2000) and L. Mandel, E. Wolf, Optical Coherence and Quantum Optics (Cambridge, NY, 1995)

[5] "Every theoretical physicist who is any good knows six or seven different theoretical representations for exactly the same physics. He knows that they are all equivalent, and that nobody is ever going to be able to decide which one is right at that level, but he keeps them in his head, hoping that they will give him different ideas for guessing." Feynman, MIT Press Character of Physical Law (1985), 168.

[6] Quantum field theory textbooks that I am aware of do not mention it. For example, the following first rate texts do not discuss Schrödinger wavefuntionals: F. Mandl and G. Shaw, *Quantum Field Theory* (2$^{nd}$ ed, Wiley, West Sussex 2010), P. Ramond, *Field Theory: A Modern Primer* (2$^{nd}$ ed, Westview Press, Boulder 2001), C. Itzykson and J. Zuber, *Quantum Field Theory* (McGraw-Hill, USA 1980), M. Peskin and D. Schroeder, *Quantum Field Theory* (Perseus Books, MA 1995), M. Kaku, *Quantum Field Theory: A Modern Introduction* (Oxford University Press, NY 1993), A. Zee, *Quantum Field Theory in a Nutshell* (Princeton Univ. Press, NJ 2003). W. Greiner and J. Reinhardt, *Field Quantization* (Springer-Verlag, Berlin, 1996). S. Weinberg, *The Quantum Theory of Fields, Volume I and II* (Cambridge University Press, NY 1995, 1996).

[7] K. Symanzik, *Schrodinger Representation and Casimir Effect in Renormalizable Quantum Field Theory*, Nucl. Phys. B 190, 1-44 (1981) showed the Schrödinger representation was renormalizable. These paper discuss it: B. Hatfield, *Quantum Field Theory of Point Particles and Strings* (Perseus, Cambrdige MA, 1992), R. Jackiw, *Analysis on Infinite-Dimensional Manifolds Schrödinger Representation for Quantized Fields in Field Theory and Particle Physics*, O. Eboli, et al. eds. (World Scientific, Brazil, 1989) 78-143p.

understanding the approach's fundamental nature and hopefully in spawning its further use.

In section 1, first the basic formalism is introduced in the process of setting up the calculation for the field structure of the single photon wave functional. Next, Parseval's theorem is introduced to analyze that structure statistically, as it must be since the quantum mechanical vector potential is fundamentally stochastic. This statistical formalism is commonly used, for example, in electrical engineering where one looks, amidst noise, for the spectral content of a signal.[8] This is fundamentally is needed here. The calculation shows that the amplitude of $A$ for the single photon of momentum $p$ is as intuitively expected sinusoidal of the frequency $v = |p|/h$.

In section 2, the formalism is generalized to the case of $n$ identical free photons; the full explicit form of the wavefunctional is given up to 4 photons and then the core form of the $n$-photon wavefunctional is given. In section 3, the multi-photon case with *different* momenta is resolved by giving the explicit results for the two photon different-momenta case. In each of these last two sections, the results prove that 1-photon sinusoidal structure is found, as intuitively expected, also in the general $n$-photon systems.

We take metric, $\eta_{\mu\nu}$, to have signature $(+,-,-,-)$ and components $\{0,1,2,3\}$ and $\hbar = 1$, and, ultimately, $c = 1$.

## Single Photon Wave-functional

Assuming one is operating in the Lorenz gauge, where the classical equations of motion are separable, we get the equations for the vector potential components in the presence of a current source:

(1) $$\frac{1}{c^2}\frac{\partial^2 A^\mu}{\partial t^2} - \nabla^2 A^\mu = \frac{4\pi}{c} J^\mu$$

We are not here interested in the *production* of the fields, so we consider only the source-less case, which can arise from considering only regions sufficiently far from the sources or, for conceptual simplicity, we can simply posit a given field of radiation as an initial condition. Furthermore, since our interest is in radiation, for which the scalar potential is not needed, we simply ignore it.[9] This leaves taking $c = 1$ and using the Einstein summation convention:

(2) $$\ddot{A}^i(x) - \partial_j^{\,2} A^i(x) = 0, \text{ where } x \text{ is the 4-vector coordinate.}$$

(this EOM is associated with $\mathcal{L} = \frac{1}{2}\partial_\mu A^i \partial^\nu A^i - \frac{1}{2}m^2 A^{i2}$, the spatial Proca Lagrangian, with $m = 0$)

---

[8] R. M. Howard, *Principles of Random Signal Analysis and Low Noise Design: The Power Spectral Density and Its Applications* (John Wiley & Sons, 2002).

[9] In this way, we sidestep problems with quantizing the full special relativistic field ($A^\mu$) in the Lorentz gauge. A more formal way to ignore the scalar potential is assume an arbitrarily small mass (rather than zero mass) in the Proca Lagrangian and then use the Lorenz gauge condition to eliminate the scalar potential (since in the massive case there are three independent degrees of freedom, rather than the two of the massless field).

Moving to second quantization operators, recalling the classical canonical momentum $\pi^i(x) \equiv \partial A^i(x)/\partial t$, this equation implies the following Hamiltonian operator in Heisenberg picture.

(3) $$\hat{H} = \frac{1}{2}\int d^3x \left( \hat{\pi}^{i\,2}(x,t) + \nabla\hat{A}^i(x,t)\cdot\nabla\hat{A}^i(x,t) \right),$$

where we impose (equal time) canonical quantization:[9]
$$[\hat{A}^i(x,t), \hat{\pi}_j(x,t)] = i\delta_{ij}\delta(x-x').$$

Thus, the Schrödinger equation for the wave functional $\Psi$ is: [10]

(4) $$i\frac{\partial \Psi(A^i)}{\partial t} = \int d^3x \frac{1}{2}\left( -\frac{\delta^2}{\delta A^{i\,2}} + (\nabla \hat{A}^i)^2 \right)\Psi(A^i),$$

Here we used:

(5) $$i\left( \dot{\hat{A}}^i = \hat{\pi}^i \to -i\frac{\delta}{\delta A^i} \right) = \frac{\delta}{\delta A^i}$$

The free solution (in the Heisenberg picture) for the vector potential operator is, taking motion along z-axis, and assuming only one polarization (thereby also limiting $A$ eigenstates, $|A\rangle$, to one polarization at a time [11]):[12,13]

(6) $$\hat{A}^{i(\lambda)}(x,t) = \frac{1}{(2\pi)^{3/2}} \int \frac{d^3p}{\sqrt{2E_p}} \left( \hat{a}_p^{(\lambda)}(t) e^{ip\cdot x} + \hat{a}_p^{\dagger(\lambda)}(t) e^{-ip\cdot x} \right) \delta^{i\lambda}$$

where: $\hat{a}_p^{(\lambda)}(t) = \hat{a}_p^{(\lambda)} e^{-iEt}$, $E_p = |p|$, $p = (0,0,p)$
$\lambda \in \{1,2\}$, (two independent polarizations)

From here on, we will usually drop the formal $t$ dependence of the creation and annihilation operators, so that by also dropping explicit polarization notation (by taking $\lambda \to i$), we can write simply $\hat{a}_p^i$. So, with $[\hat{a}_p, \hat{a}_{p'}] = [\hat{a}_p^\dagger, \hat{a}_{p'}^\dagger] = 0$, $[\hat{a}_p, \hat{a}_{p'}^\dagger] = \delta(p-p')$,

$$\hat{H} = \int E_p \left( \hat{a}_p^\dagger \hat{a}_p + \frac{1}{2} \right) d^3p.$$

Adding (6) to its time derivative and taking the Fourier transform gives, dropping the polarization index from $A$:

(7) $$\hat{a}_p^i = \frac{1}{(2\pi)^{3/2}} \frac{1}{\sqrt{2E_p}} \int \left( E_p \hat{A}^i(x,t) + i\dot{\hat{A}}^i(x,t) \right) e^{-ip\cdot x} d^3x$$

---

[10] The equation is the field theory analog, for a given spectral mode of $A$ at each point in space, of the ordinary quantum mechanical oscillator.
[11] Note this still preserves the commutation relations (since it becomes analogical to the scalar field case).
[12] The equation of motion for the vector potential operator can be shown, via the quantum generalization of Hamilton's equations, to be the analog of the classical vector potential field's equation of motion.
[13] Note the measure in the integrand is not *manifestly* relativistic invariant, the manifestly Lorentz invariant: $d^4p\,\delta(p^2-m^2)u(p_0)$ reduces, after integrating over $p_0$ and with appropriate normalization, to the 3D measure given in the text body. Note that a factor of $1/\sqrt{2E}$ has been absorbed into the definition of the creation operators to avoid a factor of $2E$ in the creation operator commutation relations.

Note: $x$ is the spatial coordinate.

Now, noting that the QFT creation operators are analogous to the creation and annihilation operators of the quantum mechanical harmonic oscillator (QMHO) ($\hat{a} \sim \hat{x} + \hat{p}$), we can apply the well known formal method of calculating the QMHO energy eigenfunctions to calculate the wave-functionals of the $n$-photon (free) $A$-field. We start with the one photon case.

In particular, using (5) in (7), we define the relevant wave functional as the probability amplitude for finding a field which is prepared in the state called a (ideal) "single photon of momentum $p$" (and described by the ket $|p\rangle = a_p^\dagger |0\rangle$) in the state $|A^i\rangle$, which is the eigenstate of $\hat{A}^i(x)$. We get, suppressing the explicit time dependence (by taking, e.g., $t = 0$):

(8) $\quad \Psi_{\bar{p}}(A^i(x)) = \langle A^i | \bar{p} \rangle = \langle A^i | 1_{\bar{p}} \rangle = \langle A^i | a_{\bar{p}}^\dagger | 0 \rangle$

$$= \frac{1}{(2\pi)^{3/2}} \frac{1}{\sqrt{2E_{\bar{p}}}} \int \left( E_{\bar{p}} A^i(x) \Psi_0(A^i(x)) - \frac{\delta}{\delta A^i(x)} \Psi_0(A^i(x)) \right) e^{i\bar{p}\cdot x} d^3 x$$

where: $\Psi_0 \equiv \Psi_0(A^i(\cdot)) \equiv \langle A^i(\cdot) | 0 \rangle$ is the vacuum field wave functional which is complex valued and not a function of $x$, the argument of $A$, despite its notational presence above (to distinguish between position and momentum field representations).

So:

(9) $\quad \Psi_{\bar{p}}(A^i(x)) = \frac{1}{\sqrt{2E_{\bar{p}}}} \left( E_{\bar{p}} \tilde{A}^i(-\bar{p}) \Psi_0(A^i(x)) - \frac{\delta \Psi_0(A^i(x))}{\delta \tilde{A}^i(\bar{p})} \right),$

where: $A^i(x) \equiv \frac{1}{(2\pi)^{3/2}} \int d^3\bar{p}\, \tilde{A}^i(\bar{p}) e^{i\bar{p}\cdot x}$, which gives: $\delta A^i(x)/\delta \tilde{A}^i(\bar{p}) = e^{i\bar{p}\cdot x}$, and allows us to take $A^i(x)$ to be an implicit functional of $\tilde{A}^i(\bar{p})$). Also: $\frac{\delta \Psi_0}{\delta \tilde{A}^i(\bar{p})} = \int \frac{\delta A^i(x)}{\delta \tilde{A}^i(\bar{p})} \frac{\delta \Psi_0(A^i(x))}{\delta A^i(x)} d^3 x$.

We need the form of the vacuum wave functional, $\Psi_0(A^i)$, which arises from noting:

(10)a $\quad \langle A^i | \hat{a}_p | 0 \rangle = 0$

And, using (5) and (7) in this equation, we get:

(10)b $\quad \int d^3 x\, e^{-i p \cdot x} \left( \frac{\delta \Psi_0(A^i(x))}{\delta A^i(x)} + E_p A^i(x) \Psi_0(A^i(x)) \right) = 0$

Using, analogous to above, $\tilde{A}^i(p) \equiv \frac{1}{(2\pi)^{3/2}} \int d^3 x\, A^i(x) e^{-i\bar{p}\cdot x}$, we can write:

(11) $$\frac{\delta \Psi_0}{\delta \tilde{A}^i(p)} + E_p \tilde{A}^i(-p) \Psi_0 = 0.$$

So, that finally, the **vacuum wave functional**: is seen to be:

(12) $$\Psi_0 = N \exp\left(-\frac{1}{2}\int d^3 p\, E_p \tilde{A}^i(-p)\tilde{A}^i(p)\right)$$

where we drop its $A^i(x)$ dependence on $\Psi_0$ because we are here displaying its implicit dependence on $\tilde{A}^i(p)$ and $N$ is the normalization of the wave functional.

Now, we can get the **wave functional of the ideal *single photon* of momentum** $\bar{p}$ by substituting (12) in (9), ignoring the normalization of the wave-functional:

(13) $$\Psi_{\bar{p}}(A^i(x)) = \sqrt{2|\bar{p}|}\tilde{A}^i(-\bar{p})\, e^{-\frac{1}{2}\int |p||A^i(p)|^2 d^3 p},$$

where we used the fact that $A(x)$ is real implies $\tilde{A}^*(p) = \tilde{A}(-p)$.

From this, we get the probability of finding the field in a state with a Fourier transform in a small region of functions space $D\tilde{A}^i$ near $\tilde{A}^i(\bar{p})$ is:

(14) $$P(\tilde{A}^i(\bar{p})) = |\Psi_{\bar{p}}(A^i)|^2 = 2|\bar{p}||\tilde{A}^i(\bar{p})|^2\, e^{-\int |p||A^i(p)|^2 d^3 p}.$$

We need to find the magnitude of $A^i(x)$, or equivalently the "energy" , $|A|^2$, that is the most likely.[14] Parseval's theorem tells us that the "energy" spectral density of a spatial function $A^i(x)$ is $D^i(p) \equiv |\tilde{A}^i(p)|^2$, so in these terms, (14) can be recast as:

(15) $$P(D^i) = |\Psi_{\bar{p}}(A)|^2 = 2|\bar{p}|\frac{(D^i(\bar{p}) + D^i(-\bar{p}))}{2}\, e^{-\int |p||D^i(p)| d^3 p}$$

where, for the multiplicative factor, we have replaced $D$ by the given sum to make explicit the reflective symmetry of $D$ in $\bar{p}$,[15,16] which we need to see the consequences of. (Note: such a replacement does not change anything in the integral in the exponent).

We need to extremize this probability distribution, so we set the first variation to zero giving:

---

[14] Note we are here interested in *A*, not *E* or *B*. We want to know the most likely vector potential structure.
[15] $D(p)$ must be even in *p* for the autocorrelation function to be positive, as it must be.
[16] Note: even though, for simplicity, we have set $p = (0,0,p)$, no generality was lost in so doing; so we may at any moment let *p* point in any direction we like.

$$(16) \quad \frac{\delta P(D(\bar{p}))}{\delta D(p)} \propto \frac{\delta}{\delta D}\left(\frac{(D(\bar{p})+D(-\bar{p}))}{2} e^{-\int |p|D d^3 p}\right)$$

$$= \frac{(\delta(p-\bar{p})+\delta(p+\bar{p}))}{2} e^{-\int |p|D d^3 p} - |p|D(p) e^{-\int |p|D d^3 p} = 0,$$

where we have dropped the component index, $i$, to simplify the notation. Thus, it is seen that the **most likely energy spectral density** is:

$$(17) \quad D_{max}(p) = \frac{(\delta(p-\bar{p})+\delta(p+\bar{p}))}{2|\bar{p}|}.$$

This result agrees with the obvious fact that the maximum of $P(D)$ occurs when the integral in the exponent is minimal and the multiplicative factor term is maximal, for at the $p = \pm\bar{p}$, the factor is infinite, while the exponent is finite.

Using (17) gives the autocorrelation function:[17]

$$(18) \quad R_{AA}(x) = \mathbf{E}[A(y)A(y+x)] = \int_{-\infty}^{\infty} D_{max}(p) e^{i p \cdot x} d^3 p = \frac{\cos \bar{p} \cdot x}{|\bar{p}|}.$$

Hence, we see the strong sinusoidal dependence is present in the stochastic $A$-field that characterizes a single (ideal) photon.

## N-Photon Wavefunctionals

Extending what we did in equation (8), and in analogy to the generation of the energy eigenstate wave functions for the harmonic oscillator in ordinary quantum mechanics,[18] we construct the wavefunctional corresponding to $n$ photons: [19]

$$(19) \quad \Psi_{n\bar{p}}(A^i(x)) = \langle A^i | n\bar{p} \rangle = \langle A^i | n_{\bar{p}} \rangle = \langle A^i | (a_{\bar{p}}^\dagger)^n | 0 \rangle,$$

Substituting (7) and generalizing from (8):

$$(20) \quad \Psi_{n\bar{p}}(A^i(x)) = \frac{1}{(2\pi)^{3n/2}} \frac{1}{(2E_p)^{n/2}} \langle A^i | \left(\int e^{i\bar{p}\cdot x}\left(E_{\bar{p}} \hat{A}^i(x) - \frac{\delta}{\delta A^i(x)}\right) d^3 x\right)^n | 0 \rangle$$

Note: we will need to use: $\dfrac{\delta \tilde{A}(p)}{\delta A(x)} = e^{-ip\cdot x}$

We have already done $n = 0$ and $n = 1$ case; we now calculate the next three cases (see Appendix A). As a result, we can write the wavefunctionals for 0,1,2, 3 and 4 photons (leaving aside complicating multiplicative factors):

$$(21) \quad \Psi_0(A^i(x)) \propto e^{-\frac{1}{2}\int |p| |\tilde{A}^i(p)|^2 d^3 p}$$

---

[17] In relation to conventional notation, we have: $D(p) \equiv S_{AA}(p) = \int_{-\infty}^{\infty} R_{AA}(x) e^{-ip\cdot x} d^3 x = \tilde{R}_{AA}(p)$.

[18] The analogy is fundamentally: $x \leftrightarrow A(x)$, $p \leftrightarrow \tilde{A}(p)$

[19] Note that the Fourier transform of the general energy eigenstate, is proportional to a state with the exact same formal structure except with x replaced by p, thus in some extended sense it is also an eigenstate of the Fourier operator.

$$(22) \quad \Psi_{\bar{p}}\left(A^i(x)\right) \propto 2|\bar{p}|\tilde{A}^i(-\bar{p})e^{-\frac{1}{2}\int |p|\left|A^i(p)\right|^2 d^3p}$$

$$(23) \quad \Psi_{2\bar{p}}\left(A^i(x)\right) \propto 2|\bar{p}|\left(2|\bar{p}|\left(\tilde{A}^i(-\bar{p})\right)^2 - \delta(2\bar{p})\right)e^{-\frac{1}{2}\int |p|\left|A^i(p)\right|^2 d^3p}$$

$$(24) \quad \Psi_{3\bar{p}}\left(A^i(x)\right) \propto 4|\bar{p}|^2\left(2|\bar{p}|\left(\tilde{A}^i(-\bar{p})\right)^3 - 3\left(\tilde{A}^i(-\bar{p})\right)\delta(2\bar{p})\right)e^{-\frac{1}{2}\int |p|\left|A^i(p)\right|^2 d^3p}$$

$$(25) \quad \Psi_{4\bar{p}}\left(A^i(x)\right) \propto 4|\bar{p}|^2 \begin{pmatrix} 4|\bar{p}|^2\left(\tilde{A}^i(-\bar{p})\right)^4 - 12|\bar{p}|\left(\tilde{A}^i(-\bar{p})\right)^2 \delta(2\bar{p}) \\ +3\left(\delta(2\bar{p})\right)^2 \end{pmatrix} e^{-\frac{1}{2}\int |p|\left|A^i(p)\right|^2 d^3p}$$

Now, we can drop the $\delta(\bar{p})$ terms, as we have obviously chosen $\bar{p} \neq 0$; so, for example, for the 4-photon case we get:

$$(26) \quad \Psi_{4\bar{p}}\left(A^i(x)\right) \propto 16|\bar{p}|^4\left(\tilde{A}^i(-\bar{p})\right)^4 e^{-\frac{1}{2}\int |p|\left|A^i(p)\right|^2 d^3p} \qquad (\bar{p} \neq 0)$$

We see the pattern. Using the formal analogy to the ordinary quantum oscillator,[20] or induction, we can write the *first try* at the solution as, dropping the component superscript, $i$, and the multiplicative factors:

$$(27) \quad \text{1st try: } \Psi_{n\bar{p}}(A(x)) \propto (-1)^n e^{-\frac{1}{2}\int |p|\left|A^i(p)\right|^2 d^3p} e^{\int |p|\left(\tilde{A}(-p)\right)^2 d^3\bar{p}} \frac{\delta^n}{\delta^n \tilde{A}(-p)} e^{-\int |p|\left(\tilde{A}(-p)\right)^2 d^3\bar{p}}.$$

This first try, however, includes the terms that correspond to the delta functions which, as just stated, we don't need; so, we write:

$$(28)\ \Psi_{n\bar{p}}(A(x)) \propto e^{-\frac{1}{2}\int |p|\left|A^i(p)\right|^2 d^3p} (-1)^n \left(-2|\bar{p}|\tilde{A}(-\bar{p})\right)^n = \left(2|\bar{p}|\right)^n \left(\tilde{A}(-\bar{p})\right)^n e^{-\frac{1}{2}\int |p|\left|A^i(p)\right|^2 d^3p}.$$

This means the corresponding probability distribution is, dropping constant multiplicative factors:

$$(29) \quad P_{n\bar{p}}(D) \propto \left|\tilde{A}(\bar{p})\right|^{2n} e^{-\int |p|\left|A^i(p)\right|^2 d^3p} = \bar{D}(\bar{p})^n e^{-\int |p|\left|A^i(p)\right|^2 d^3p},$$

where we introduce $\bar{D}(p) \equiv (D(p) + D(-p))/2$ to make $D$'s symmetry in $p$ explicit for the minimization differentiation.

Minimizing gives:

$$(30) \quad \frac{\delta P_{n\bar{p}}(D(\bar{p}))}{\delta D(p)} \propto \frac{\delta}{\delta D}\left(\left(\frac{D(\bar{p}) + D(-\bar{p})}{2}\right)^n e^{-\int |\bar{p}|\bar{D}(p) d^3p}\right)$$

$$= n\bar{D}(\bar{p})^{n-1}\left(\frac{\delta(p-\bar{p}) + \delta(p+\bar{p})}{2}\right)^n e^{-\int |\bar{p}|\bar{D}(p) d^3p} - |\bar{p}|D(\bar{p})^n e^{-\int |\bar{p}|\bar{D}(p) d^3p} = 0$$

which implies, dropping the explicit separation of $D$ in the first term:

---

[20] The Fourier transform of the energy eigenstates ($\psi_n = \exp(-x^2/2)H_n(x)$) in ordinary quantum mechanics, that we use here by analogy is: $\psi_n(p) = (i)^n \exp(p^2/2) d^n\left(\exp(-p^2)\right)/dp^n$. Note: $H_n(x) = (-1)^n \exp(x^2)d^n \exp(-x^2)/dx^n$.

$$(31) \quad nD(\bar{p})^{n-1}\left(\frac{\delta(p-\bar{p})+\delta(p+\bar{p})}{2}\right)=|\bar{p}|D(\bar{p})^n$$

So we have:

$$(32) \quad D(\bar{p})=n\left(\frac{\delta(p-\bar{p})+\delta(p+\bar{p})}{2|\bar{p}|}\right)$$

So $n$ photons have $n$ times the energy spectral density as a single one, but the same concentration of that energy into a sinusoidal form. This is a very nice result that agrees with ones intuition; one can think of the field of a photon as mostly having sine-like behavior, but on top of a stochastic background. (Of course, more information can be extracted from the wavefunctional, such as the moments of the distribution.)

### Many-photon Wave-functionals with Different Momenta

The above formalism is generalizable to the photons of different momenta. The basic principle can be illustrated by the two photon case, one with momentum $p_1$ the other with momentum $p_2$.

We can extend the general formalism to this case by applying the creation operator twice $|p_1,p_2\rangle = a^\dagger(p_2)a^\dagger(p_1)|0\rangle$); we get, temporarily dropping the factor $e^{-\int |\bar{p}|\bar{D}(p)d^3p}$:

$$(33) \quad P_{p_1,p_2}=\psi_{p_1,p_2}\psi^*_{p_1,p_2}\propto \begin{pmatrix}(4|p_1||p_2|A(p_1)A(p_2)-2|p_1|\delta(p_1+p_2))\times\\(4|p_1||p_2|A^*(p_1)A^*(p_2)-2|p_1|\delta(p_1+p_2))\end{pmatrix}$$

$$=8|p_1||p_2|(2|p_1||p_2|D_1D_2-\delta(p_1+p_2)(A_1^*A_2^*+A_1A_2))+4|p_1|^2(\delta(p_1+p_2))^2$$

First, we consider the most general and most interesting case (for these idealized photons), which is also the simplest to calculate.

Case I: $p_1 \neq -p_2$.

This means the delta functions in (33) are not in play, so we have, in terms of the previously defined $\bar{D}$ and reinserting the exponential term:

$$(34) \quad P_{p_1,p_2}(D(p_1),D(p_2))\propto 16|p_1|^2|p_2|^2\bar{D}(p_1)\bar{D}(p_2), e^{-\int |\bar{p}|\bar{D}(p)d^3p}$$

We want: $\dfrac{\delta P_{p_1,p_2}}{\delta D_1}=0$ and $\dfrac{\delta P_{p_1,p_2}}{\delta D_2}=0$.

Since this is formally same as calculation previously done, we get:

$D_i=\dfrac{\delta(p-p_i)+\delta(p-p_i)}{2|p_i|}$. So, as expected, we get one field structure with strong sinusoidal structure of frequency associated with $p_1$ and one associated with $p_2$. Note equation (34) and this last result reduce to the same result for the case in which $p\equiv p_1=p_2$; to see the latter, compare equations (34) and (29). This again verifies our intuition of two photon systems.

We now move to the second, more limited case.

Case II: $p_1=-p_2$

Given our analysis in the previous sections, we would expect the two photons might cancel each other out in our (idealized photon) case. To do the calculation, we take the following definitions: $\boldsymbol{p} \equiv \boldsymbol{p}_1 = -\boldsymbol{p}_2$, $D \equiv D_i \equiv |A_i|^2 = A(-\boldsymbol{p}_i)A(\boldsymbol{p}_i) = |A(\boldsymbol{p})|^2$ and recall $A^*(\boldsymbol{p}) = A(-\boldsymbol{p})$.

Equation (33) becomes in the limit that $\boldsymbol{p}_1 \to -\boldsymbol{p}_2$:

$$(35)\ P_{p_1,p_2}(D_1,D_2) \propto 8|\boldsymbol{p}|^2 \left(2|\boldsymbol{p}|^2 D^2 - \delta(0)\big(A(-\boldsymbol{p})A(\boldsymbol{p}) - A(\boldsymbol{p})A(-\boldsymbol{p})\big)\right) + 4|\boldsymbol{p}|^2 \big(\delta(0)\big)^2$$

When finding the maximum, it is the highest order delta functions that matter. Hence, only the last term needs to be considered (the second term cannot contribute a second order delta function because the piece of that term in parenthesis goes to zero) so we have, recalling the exponential term:

$$P_{p_1,p_2}(D_1,D_2) \propto 4|\boldsymbol{p}|^2 \big(\delta(\boldsymbol{p}_1+\boldsymbol{p}_2)\big)^2 e^{-\int |\boldsymbol{p}| D(\boldsymbol{p}) d^3 p} \quad (\boldsymbol{p} \equiv \boldsymbol{p}_1 = -\boldsymbol{p}_2)$$

Finding the extremum, we get:

$$\frac{\delta P_{p_1,p_2}}{\delta D_i} \propto \big(\delta(\boldsymbol{p}_1+\boldsymbol{p}_2)\big)^2 \frac{\delta}{\delta D_i} e^{-\int |\boldsymbol{p}| D(\boldsymbol{p}) d^3 p} = -\big(\delta(\boldsymbol{p}_1+\boldsymbol{p}_2)\big)^2 |\boldsymbol{p}| D(\boldsymbol{p}) e^{-\int |\boldsymbol{p}| D(\boldsymbol{p}) d^3 p} = 0$$

So, $D(\boldsymbol{p}) = 0$, which means that there is no average energy in this mode in the most likely *A* amplitude distribution.

## Conclusion

We have here given the structure of the photon vector potential distribution and theoretically derived, for the first time, its key feature, thus enlightening long held, though often confused, intuitions about photons.

Like all references to quantum mechanical states, the term photon refers to a system statistically, not individually; quantum mechanics only predicts probable outcomes, so needs many experiments on different systems (experiments on an "ensemble" of systems) in the same state (or approximately the same state) to verify the prediction. This paper proves that a photonic state is a state which has a stochastic vector potential (*A*) filling space whose most likely amplitude (magnitude) structure (across the ensemble of systems) has a sinewave component of frequency *p / h* for each photon of momentum *p* in the system. Getting this insight into the free field states, being fundamental, can build intuition into other more complicated (e.g., interactional) QFT problems.

Recognizing the stochastic character of the QFT states can lead one to introduce Parseval's theorem (and other statistical theorems) which along with the Schrödinger formalism are intuitive tools and are required to get these results. The QFT analog of the Schrödinger equation allows one to carry all the intuition one gained in ordinary quantum mechanics about waves and particles over to QFT, including into Yang Mill's theory. It has been used in attempts to use QFT in general relativity in a 3+1 space time setting, which also points to its draw back that it is not explicitly relativistically invariant.[21] It is also required and used in interesting ways in De Broglie-Bohm (dBB) formalism of QFT.[22] Though this work fills a lacunae and is foundational (further demonstrating the nature of the fundamental entity "photon"), it requires the relatively infrequently used QFT Schrödinger formalism and this seems to be part of why it has been overlooked till now. Hopefully, this filling of the lacunae will help spur the use of the Schrödinger formalism and thereby make available the attending insights offered by its distinct perspective on QFT.[23,24]

---

[21] D.V. Long, G.M. Shore, *The Schrödinger Wave Functional and Vacuum States in Curved Spacetime*, Nucl. Phys. B 530, 247-278 (1998) D.V. Long, G.M. Shore, *The Schrodinger Wave Functional and Vacuum States in Curved Spacetime II: Boundaries and Foliations*, Nucl. Phys. B 530, 279-303 (1998). P.R. Holland, The De Broglie-Bohm Theory of Motion and Quantum Field Theory, Phys. Rept. **224** No. 3,

# Appendix A
### Calculation of the wave-functionals for 2, 3 and 4 photon systems.

(36) $$\Psi_{n\bar{p}}\left(A^i(x)\right) = \langle A^i | n\bar{p}\rangle = \langle A^i | n_{\bar{p}}\rangle = \langle A^i | \left(a_{\bar{p}}^\dagger\right)^n | 0 \rangle,$$

Substituting (7) and generalizing from (8):

(37) $$\Psi_{n\bar{p}}\left(A^i(x)\right) = \frac{1}{(2\pi)^{3n/2}} \frac{1}{(2E_p)^{n/2}} \langle A^i | \left( \int e^{i\bar{p}\cdot x}\left(E_{\bar{p}}\hat{A}^i(x) - \frac{\delta}{\delta A^i(x)}\right)d^3x \right)^n | 0 \rangle$$

Note: we will need to use: $\dfrac{\delta \tilde{A}(\bar{p})}{\delta A(x)} = e^{-i\bar{p}\cdot x}$ (and perhaps $\dfrac{\delta A(x)}{\delta \tilde{A}(\bar{p})} = e^{i\bar{p}\cdot x}$ )

Dropping multiplicative factors to reduce space and focus on form of the equations, we can write the generic equation that we will use below:

(38) $$\Psi_{n\bar{p}}\left(A^i(x)\right) = = \left( |\bar{p}| \tilde{A}^i(-\bar{p}) - \frac{\delta}{\delta \tilde{A}^i(\bar{p})} \right) \Psi_{(n-1)p}\left(A^i(x)\right)$$

Using this and recalling $\Psi_{\bar{p}}\left(A^i(x)\right) = 2|\bar{p}|\tilde{A}^i(-\bar{p}) e^{-\frac{1}{2}\int |p| |A^i(p)|^2 d^3 p}$, we now calculate the 2, 3 and 4 photon wavefunctionals.

**n=2** case :

---

(39)
$$\Psi_{2\bar{p}}\left(A^i(x)\right) = \left(|\bar{p}|\tilde{A}^i(-\bar{p}) - \frac{\delta}{\delta\tilde{A}^i(\bar{p})}\right)\left(|\bar{p}|\tilde{A}^i(-\bar{p}) - \frac{\delta}{\delta\tilde{A}^i(\bar{p})}\right)\Psi_0\left(A^i(x)\right)$$

$$= \left(|\bar{p}|\tilde{A}^i(-\bar{p}) - \frac{\delta}{\delta\tilde{A}^i(\bar{p})}\right)\Psi_{\bar{p}}\left(A^i(x)\right) = |\bar{p}|\tilde{A}^i(-p)\Psi_{\bar{p}}\left(A^i(x)\right) - \frac{\delta\Psi_{\bar{p}}\left(A^i(x)\right)}{\delta\tilde{A}^i(\bar{p})}$$

$$= 2|\bar{p}|\left(|\bar{p}|\tilde{A}^i(-\bar{p})\tilde{A}^i(-\bar{p})e^{-\frac{1}{2}\int|p||A^i(p)|^2 d^3p} - \frac{\delta\left(\tilde{A}^i(-\bar{p})e^{-\frac{1}{2}\int|p||A^i(p)|^2 d^3p}\right)}{\delta\tilde{A}^i(\bar{p})}\right)$$

$$\Psi_{2\bar{p}}\left(A^i(x)\right) = 2|\bar{p}|\left(2|\bar{p}|\left(\tilde{A}^i(-\bar{p})\right)^2 - \delta(2\bar{p})\right)e^{-\frac{1}{2}\int|p||A^i(p)|^2 d^3p}$$

**n=3** case:
(40)
$$\Psi_{3\bar{p}}\left(A^i(x)\right) = \left(|\bar{p}|\tilde{A}^i(-\bar{p}) - \frac{\delta}{\delta\tilde{A}^i(\bar{p})}\right)\Psi_{2\bar{p}}\left(A^i(x)\right)$$

$$= 2|\bar{p}|\left(|\bar{p}|\tilde{A}^i(-\bar{p}) - \frac{\delta}{\delta\tilde{A}^i(\bar{p})}\right)\left(2|\bar{p}|\left(\tilde{A}^i(-\bar{p})\right)^2 - \delta(2\bar{p})\right)e^{-\frac{1}{2}\int|p||A^i(p)|^2 d^3p}$$

$$= 2|\bar{p}|\left(\begin{array}{c}\left(2|\bar{p}|^2\left(\tilde{A}^i(-\bar{p})\right)^3 - |\bar{p}|\tilde{A}^i(-p)\delta(2\bar{p})\right) \\ -\left(4|\bar{p}|\tilde{A}^i(-\bar{p})\delta(2\bar{p}) - |\bar{p}|\left(2|\bar{p}|\left(\tilde{A}^i(-\bar{p})\right)^3 - \tilde{A}^i(-\bar{p})\delta(2\bar{p})\right)\right)\end{array}\right)e^{-\frac{1}{2}\int|p||A^i(p)|^2 d^3p}$$

So,
$$\Psi_{3\bar{p}}\left(A^i(x)\right) = 4|\bar{p}|^2\left(2|\bar{p}|\left(\tilde{A}^i(-\bar{p})\right)^3 - 3\left(\tilde{A}^i(-\bar{p})\right)\delta(2\bar{p})\right)e^{-\frac{1}{2}\int|p||A^i(p)|^2 d^3p}$$

**n=4** case:

(41)
$$\Psi_{4\bar{p}}\left(A^i(x)\right) = \left(|\bar{p}|\tilde{A}^i(-\bar{p}) - \frac{\delta}{\delta\tilde{A}^i(\bar{p})}\right)\Psi_{3\bar{p}}\left(A^i(x)\right)$$

$$= 4|\bar{p}|^2 \begin{pmatrix} |\bar{p}|\tilde{A}^i(-\bar{p})\left(2|\bar{p}|(\tilde{A}^i(-\bar{p}))^3 - 3(\tilde{A}^i(-\bar{p}))\delta(2\bar{p})\right)e^{-\frac{1}{2}\int|p||A^i(p)|^2 d^3p} \\ -\dfrac{\delta\left(\left(2|\bar{p}|(\tilde{A}^i(-\bar{p}))^3 - 3(\tilde{A}^i(-\bar{p}))\delta(2\bar{p})\right)e^{-\frac{1}{2}\int|p||A^i(p)|^2 d^3p}\right)}{\delta \tilde{A}^i(\bar{p})} \end{pmatrix}$$

$$= 4|\bar{p}|^2 \begin{pmatrix} 2|\bar{p}|^2(\tilde{A}^i(-\bar{p}))^4 - 3|\bar{p}|(\tilde{A}^i(-\bar{p}))^2 \delta(2\bar{p}) \\ -\left(6|\bar{p}|(\tilde{A}^i(-\bar{p}))^2 \delta(2\bar{p}) - 3(\delta(2\bar{p}))^2 - |\bar{p}|\tilde{A}^i(-\bar{p})\left(2|\bar{p}|(\tilde{A}^i(-\bar{p}))^3 - 3(\tilde{A}^i(-\bar{p}))\delta(2\bar{p})\right)\right) \end{pmatrix} e^{-\frac{1}{2}\int|p||A^i(p)|^2 d^3p}$$

So,

$$\Psi_{4\bar{p}}(A^i(x)) = 4|\bar{p}|^2 \left(4|\bar{p}|^2(\tilde{A}^i(-\bar{p}))^4 - 12|\bar{p}|(\tilde{A}^i(-\bar{p}))^2 \delta(2\bar{p}) + 3(\delta(2\bar{p}))^2\right) e^{-\frac{1}{2}\int|p||A^i(p)|^2 d^3p}$$

Recalling that finally we take $\bar{p} \neq 0$, we first drop the squared deltas:

$$\Psi_{4\bar{p}}(A^i(x)) = 16|\bar{p}|^3 \left(|\bar{p}|(\tilde{A}^i(-\bar{p}))^4 - 3(\tilde{A}^i(-\bar{p}))^2 \delta(2\bar{p})\right) e^{-\frac{1}{2}\int|p||A^i(p)|^2 d^3p}$$

Ignoring the delta function completely gives:

$$\Psi_{4\bar{p}}(A^i(x)) = 16|\bar{p}|^4 (\tilde{A}^i(-\bar{p}))^4 \, e^{-\frac{1}{2}\int|p||A^i(p)|^2 d^3p}$$